| | |
|---|---|
| Title | **Effect of limiting the cathode surface on direct current microhollow cathode discharge in helium** |
| Authors | T. Dufour[1], R. Dussart[1] P. Lefaucheux[1] P. Ranson[1] L. J. Overzet[2] M. Mandra[2], J.-B. Lee[2] and M. Goeckner[2] |
| Affiliations | [1]Groupe de Recherches sur l'Energétique des Milieux Ionisés, Université d'Orléans, Orléans 45067, France <br> [2]University of Texas at Dallas, Richardson, Texas 75083, USA |
| Ref. | Applied Physics Letters, 2008, Vol. 93, Issue 7, 071508 |
| DOI | http://dx.doi.org/10.1063/1.2966144 |
| Abstract | This paper describes how to light several microdischarges in parallel without having to individually ballast each one. The V-I curve of a microhollow cathode discharge is characterized by a constant voltage in the normal glow regime because the plasma is able to spread over the cathode surface area to provide the additional secondary electrons needed. If one limits the cathode surface area, the V-I characteristic can be forced into an abnormal glow regime in which the operating voltage must increase with the current. It is then possible to light several microdischarges mounted in parallel without ballasting them individually. |

# I. Introduction

Microdischarges are nonequilibrium discharges spatially confined to dimensions smaller than 1 mm. They are a promising approach to the generation and maintenance of stable dc glow discharges at atmospheric pressure and they present interesting challenges for plasma science (impact of quantum electrodynamics on spontaneous emission rate and quasineutrality breaking).[1] Parallel microplasmas can be created using several kinds of microdevices.[2] Some teams use ballasts to individually initiate the plasma in each microcavity independently.[3,4]

The aim of this work is to investigate and optimize the ignition of parallel microdischarges without individual ballasts to create an array. This area can be extended by the way of a third plane electrode[5,6] or a nozzle electrode.[7] The microhollow cathode (MHC) used in this experiment is a Ni:Al$_2$O$_3$:Ni sandwich structure having one or several holes of 250 m in diameter. The dielectric layer (Al$_2$O$_3$) has a thickness as high as 250 µm whereas the nickel electrode thickness is only 6 µm. Ni is deposited by electrodeposition process. The holes are made by laser drilling.

# II. Experimental setup

The MHC is installed inside a vacuum chamber linked to a primary pump. The discharge can operate in a range of pressures from 100 to 1000 Torr. Experiments were carried out in He. A 0–2500 V dc (300 W) power supply is used to operate the microdischarges. Electrical measurements are performed using a Tektronix oscilloscope and high voltage probes. The discharge voltage is measured between input and output resistances (R$_1$=19 kOhm, 16 W and R$_2$=1 kOhm) and the discharge current through R2 is connected to the ground. An imaging system including an intensified charge coupled device (ICCD) camera (512*512 Princeton Instruments) and an optical system for magnification was installed in front of the microcavity.

# III. Results & Discussion

V-I curves for a single hole microdevice for four pressures are presented in Fig. 1. These characteristics were obtained by measuring the current during linear increase and decrease in supply voltage (40 s period). Up until now, experiments carried out by other teams[8–11] with the same MHC configuration have shown the existence of an abnormal glow regime for small values of the discharge current.[12] The effect of the cathode thickness has been predicted by a model[12] based on solutions of fluid equations in the drift-diffusion approximation for the electron and ion transport coupled with Poisson's equation. If the discharge is sustained by electron emission from the outer surface of the cathode (case corresponding to a thin cathode at low current), the V-I curve has a negative slope before showing a constant voltage versus current. If the discharge is sustained by electron emission from inside the hollow cathode (case corresponding to a thick cathode at low current), the V-I curve has a positive slope, which corresponds to the abnormal glow regime. In our case, the cathode thickness being very small, we observe a slight decrease in the voltage at a very low current as predicted by the model[12] (only for decreasing values of supply voltage) before the normal glow regime (inset of Fig. 1). Other teams usually use much thicker cathodes (about 100 µm).[8–11] As a consequence, they have observed a positive V-I slope at a low current in good agreement with the same model. The breakdown voltage values have a Paschen law behavior as also observed by other teams.[9]

After breakdown, we observed a self-pulsing regime of the microdischarge, which was studied in detail by Rousseau and Aubert.[10] During increasing current, the discharge is no longer confined inside the cavity. It extends to the outer cathode surface to cover a large area. Thus, electron and ion production is increased by sheath expansion. We obtain a roughly constant discharge voltage resulting from the equilibrium between the size of the discharge spread and the value of the discharge current. This is the normal glow regime. As part of decreasing the supply voltage, a hysteresis is observed in the transition between the breakdown potential and the normal glow regime as seen in a typical dc discharge at low pressure.[13]

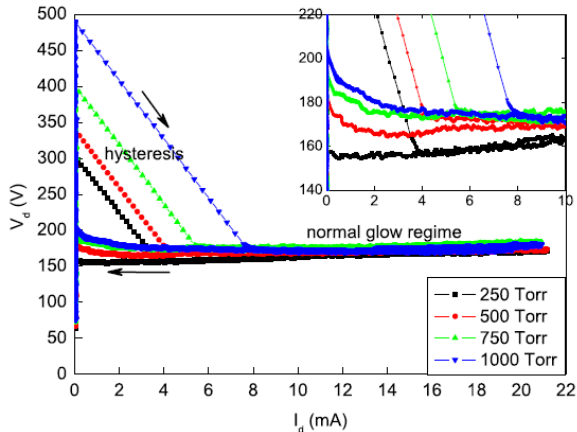

*FIG. 1. V-I curves of a 250 µm single hole MHC device in helium. (Inset) Zoom of the curves at low current.*

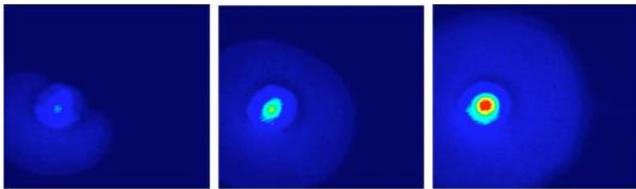

*FIG. 2. Microdischarge spread on the cathode surface for discharge currents of 5.5, 10.3, and 19.3 mA, 750 Torr, He.*

The big difference between the breakdown voltage and the operating voltage can be an issue to operate a multiple hole device without ballasting each of the microdischarges individually. Indeed, in this case, once a first hole has ignited, the voltage between the anode and cathode drops to a lower value well below the breakdown potential, which therefore makes the second cavity hard to breakdown independently. However, by increasing the discharge current in the normal glow regime, the microdischarge spreads out of its cavity, as shown by the optical emission intensity in Fig. 2 and can potentially initiate microdischarges in the nearest holes. However, in our experiment, nearest cavities do not necessarily initiate in priority. Another parameter influences the ignition of some holes. Due to the laser drilling process, holes are not exactly all the same at the microscopic scale. Some of them have a more or less rough edge which can facilitate their ignition.

According to previous works,[4,14,15] microdischarge arrays only operate in the abnormal glow regime. In the present simply fabricated MHC devices, we do not obtain this regime even at quite high currents (20 mA). We propose to spatially limit the area of the cathode by a dielectric layer while keeping the same MHC configuration. For this purpose, we covered the cathode surface with a 50 µm thick layer of Kapton. We made a controlled circular opening through this layer in order to expose a small part of the cathode surface to ion bombardment. The circular opening around the 250 µm diameter cavity was approximately 600 µm in diameter. The resulting V-I curves for 250 and 750 Torr are shown in Fig. 3. We obtained about the same breakdown voltages as in the case without limiting the cathode surface. However, the discharge voltage $V_d$ now increases with the current showing a positive differential resistance similar to the abnormal glow of low pressure plasmas.[13] Moreover, an anticlockwise hysteresis is superposed to this abnormal glow regime.

We expect that energetic ion bombardment occurs only on the exposed cathode surface. Consequently, the amount of discharge spread can be limited by adjusting the dimension of the dielectric opening. Tests were carried out on the same single hole microdevice at atmospheric pressure for cathode opening diameters between 500 and 2500 µm. V-I curves corresponding to this study are presented in Fig. 4. The V-I curve for the smallest opening has the most pronounced slope. The reason is attributed to the limited plasma expansion.

Since the emission of secondary electrons cannot be increased by plasma expansion, $I_d$ can only be increased by an increase in the cathode fall. In contrast, for the largest opening, the microdischarge can expand. Since the cathode expansion is not limited the cathode fall voltage does not need to increase and the discharge voltage remains approximately constant, as observed in Fig. 4 for a 2500 µm opening. To summarize, the smaller the exposed cathode surface, the more pronounced the abnormal glow regime.

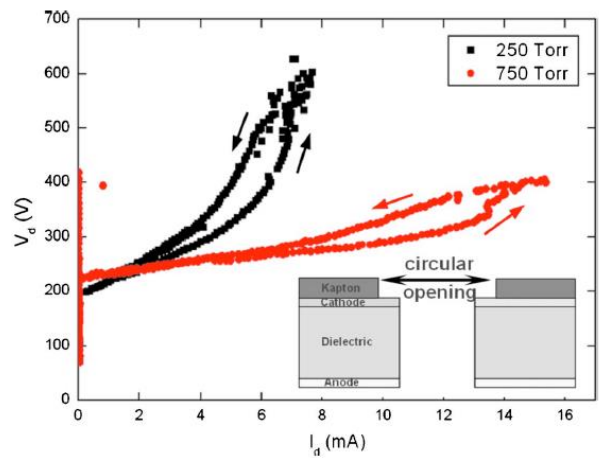

*FIG. 3. V-I curves for a single hole MHC with the cathode area limited 600*600 µm² at 250 and 750 Torr.*

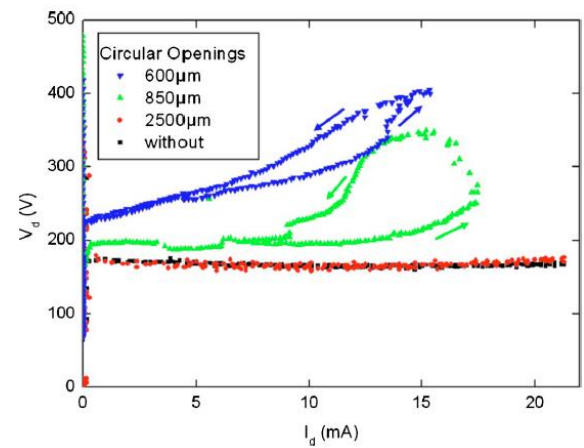

*FIG. 4. V-I curves for a single hole MHC at 750 Torr for several cathode areas.*

Let us now consider the case of a microdevice with seven holes in a line, separated center to center by a distance of 260 µm. If one does not limit the cathode, the ignition of this microdevice always begins by the ignition of a first and single microcavity for small values of the discharge current (<1 mA). When increasing $I_d$, the first microdischarge spreads on the cathode surface.[16] The spread can overlap to other holes but does not always induce their ignition. In our case, only the first hole ever ignited, as illustrated in the left inset of Fig. 5, even for large discharge currents (25 mA).

By covering the cathode surface with a dielectric layer, we limited the cathode area to 600*2700 µm$^2$. Note that no dielectric separates the seven holes. While the corresponding V-I characteristic had a larger breakdown potential (245 V) than before (185 V), it still corresponded to the ignition of the first hole. Then, for increasing discharge currents, the V-I curve shown in Fig. 5 has a positive differential resistance, corresponding to an abnormal glow regime. By limiting the cathode surface, we were able to ignite the other cavities at 27 mA without having to ballast them individually.

This curve is also characterized by small jumps in $V_d$, with amplitudes around 5 V at 9, 13, and 19 mA. According to the ICCD pictures, each jump corresponds to the ignition of an additional cavity. As the three jumps mentioned above appear below 245 V, these cavities were ignited at voltages lower than the initial breakdown potential (above 185 V). In the inset of Fig. 5 (cathode limited), we can observe that the emission of each hole is not the same, indicating that the current is not equally shared between the cavities. Once multiple holes are ignited, the system is then at a point of having less current in each cavity but more overall. As a consequence, $V_d$ is reduced.

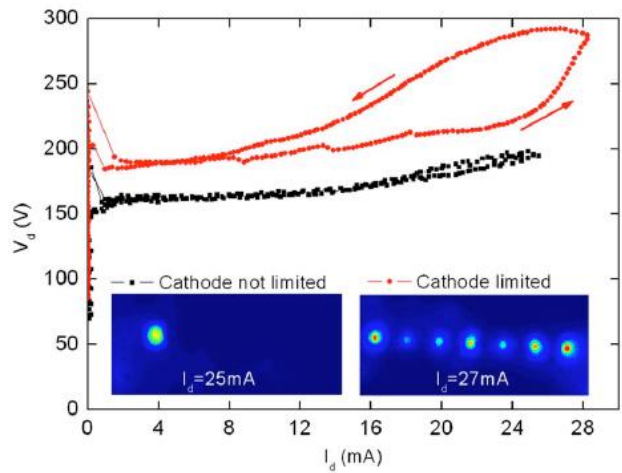

FIG. 5. V-I curve at 750 Torr for seven holes in line microdevice with/without cathode surface limited.

It would be natural to assume that once a first discharge is initiated, the breakdown potential for adjacent holes is lowered due to the seed electrons provided by the first plasma. However, there is a problem with this assumption. While these electrons could assist the ignition of an additional microdischarge, it should favor adjacent holes over further removed holes. One can clearly see in the inset picture that this is not the case. The four brightest holes are not adjacent and the ICCD pictures clearly show that it is not just the adjacent holes that ignite. Again, this effect is probably due to the fact that ignition is also influenced by the micrometric geometry of the electrodes, which are not exactly the same from a hole to another.

The hysteresis we observed in the cathode limited case seems to be due to the heating of the microdevice. Indeed, in our experiments, the period of the power supply voltage lasts by 40 s. By significantly reducing this period to less than 1 s, the hysteresis became much weaker. Long life helium metastables might also play a role in the hysteresis phenomenon; this hypothesis would need a dedicated study.

## IV. Conclusion

As a conclusion, we obtain an abnormal glow regime for high values of discharge current by limiting the outer cathode surface with a dielectric layer. The smallest cathode surface exposed to ion bombardment is required to get the most significant abnormal glow regime. This property is very helpful to initiate microdischarge arrays without individual ballasts.

## V. References

[1] R. Foest, M. Schmidt, and K. Becker, Int. J. Mass. Spectrom. 248, 87 (2006).
[2] K. H. Becker, K. H. Schoenbach, and J. G. Eden, J. Phys. D 39, R55 (2006).
[3] P. von Allmen, D. J. Sadler, C. Jensen, N. P. Ostrom, S. T. McCain, B. A. Vojak, and J. G. Eden, Appl. Phys. Lett. 82, 4447 (2003).
[4] P. von Allmen, S. T. McCain, N. P. Ostrom, B. A. Jovak, J. G. Eden, F. Zenhausern, C. Jensen, and M. Oliver, Appl. Phys. Lett. 82, 2562 (2003).
[5] R. H. Stark and K. H. Schoenbach, Appl. Phys. Lett. 74, 3770 (1999).
[6] A.-A. H. Mohamed, R. Block, and K. H. Schoenbach, IEEE Trans. Plasma Sci. 30, 182 (2002).
[7] T. Yokoyama, S. Hamada, S. Ibuka, K. Yasuoka, and S. Ishii, J. Phys. D 38, 1684 (2005).
[8] S.-J. Park, K. S. Kim, and J. G. Eden, Appl. Phys. Lett. 86, 221501 (2005).
[9] X. Aubert, G. Bauville, J. Guillon, B. Lacour, V. Puech, and A. Rousseau, Plasma Sources Sci. Technol. 16, 23 (2007).
[10] A. Rousseau and X. Aubert, J. Phys. D 39, 1619 (2006).
[11] S.-J. Park, J. G. Eden, J. Chen, and C. Liu, Appl. Phys. Lett. 85, 4869 (2004).
[12] J. P. Boeuf, L. C. Pitchford, and K. H. Schoenbach, Appl. Phys. Lett. 86, 071501 (2005).
[13] M. A. Lieberman and A. J. Lichtenberg, Principles of Plasma Discharges
and Materials Processing, 2nd ed. (Wiley, Hoboken, NJ, 2005), pp. 538– 539.
[14] S.-J. Park, J. Chen, C. Liu, and J. G. Eden, Electron. Lett. 37, 171 (2001).

[15] J. Chen, S.-J. Park, J. G. Eden, and C. Liu, J. Microelectromech. Syst. 11, 536 (2002).
[16] T. Dufour, R. Dussart, M. Mandra, and L. J. Overzet, 18th International Symposium on Plasma Chemistry, ISPC, 2007 (unpublished).